\documentclass[12pt]{article}

\newcommand{\eqn}[1]{(\ref{#1})}

\newcommand{\ft}[2]{{\textstyle{\frac{#1}{#2}}}}

\newcommand{\dr}{\raise.3ex\hbox{$\stackrel{\leftarrow}{\partial }$}{}}
\newcommand{\delr}{\raise.3ex\hbox{$\stackrel{\leftarrow}{\delta }$}{}}

\begin{document}
\renewcommand{\theequation}{\thesection.\arabic{equation}}
\csname @addtoreset\endcsname{equation}{section}
\newcommand{\dkt}{\delta_\kappa \theta}
\newcommand{\alphamsbar}{{\alpha_{\overline{\text{MS}}}}}
\newcommand{\be}{\begin{equation}}
\newcommand{\ee}{\end{equation}}
\newcommand{\bea}{\begin{eqnarray}}
\newcommand{\eea}{\end{eqnarray}}
\newcommand{\eps}{\epsilon_{\mu \nu \rho \sigma} k_1^\mu k_2^\nu k_3^\rho
k_4^\s
igma}
\newcommand{\ul}{\underline}
\newcommand{\Dsl}[1]{\slash\hskip -0.20 cm #1}
\newcommand{\al}{\alpha}
\newcommand{\s}{\sigma}
\newcommand{\w}{\omega}
\newcommand{\R}{\rho}
\newcommand{\La}{\Lambda}
\newcommand{\ens}{e^{-n\sigma}}
\newcommand{\eens}{e^{-n\sigma_0}}
\newcommand{\ch}{{\tt (*)}\ }
\renewcommand{\thefootnote}{\fnsymbol{footnote}}
\renewcommand{\thanks}[1]{\footnote{#1}} 
\newcommand{\starttext}{
\setcounter{footnote}{0}
\renewcommand{\thefootnote}{\arabic{footnote}}}
\newcommand{\N}{{\cal N}}
\newcommand{\<}{\langle}
\renewcommand{\>}{\rangle}


\begin{titlepage}
\begin{flushright}
SU-ITP-98/23 \\ KUL-TF-98/20\\ hep-th/9804099\\
\today
\end{flushright}
\vspace{.5cm}
\begin{center}
\baselineskip=16pt
{\Large\bf CONFORMAL SYMMETRY  \\

\

OF SUPERGRAVITIES IN AdS  SPACES}
\vskip 0.3cm
{\large {\sl }}
\vskip 10.mm
{\bf  Renata
Kallosh$^{a}$
and {}~Antoine Van Proeyen$^{b,\dagger}$ } \\
\vskip 0.8 cm

$^a$
 Physics Department, \\ Stanford
University, Stanford, CA 94305-4060, USA

 $^b$
Instituut voor theoretische fysica, \\ Katholieke
Universiteit Leuven, B-3001 Leuven, Belgium

 \vspace{6pt}
\end{center}
\vfill
\par
\begin{center}
{\bf ABSTRACT}
\end{center}
\begin{quote}
We show that the
background field method applied to supergravity in adS space-time provides the
path integral for the theory in the bulk
with conformal symmetry associated with the isometry of the adS space.
 This in turn allows to establish  the rigid conformal invariance of the
generating functional for
the supergravity correlators on the boundary.

\vfill
 \hrule width 5.cm
\vskip 2.mm
{\small
\noindent $^\dagger$ Onderzoeksdirecteur FWO, Belgium }
\end{quote}
\end{titlepage}

\section{Introduction}
In this note we present an observation that the effective gauge-fixed
supergravity actions have conformal (and R-symmetry) {\it in the bulk and at
the boundary} in cases when the
$adS_{p+2}
\times S^{d-p-2}$ background (with form-fields) defines a vacuum of the theory.
This  sheds a light on the  status of the recent conjecture
\cite{maldacena} about CFT/adS correspondence between the large $N$ limit of
the
$SU(N)$ Yang-Mills theory and supergravity. In particular,  we will find out
{\it what part
of the Maldacena's conjecture} \cite{maldacena}  {\it can be proved  and what
part remains to be
studied}.

Our observation is based on the existence of the background functional method,
first introduced for the gravitational field theories by Bryce De Witt
\cite{DeWitt} and extended later for the case of supergravities in
\cite{Kallosh78}. We use here the existence of the so-called background
covariant gauges in which the effective action of the theory as well as the
generating functional of the Green functions are background invariant.

In fact we are using here precisely the same idea which has allowed us to
construct a conformal theory of branes in \cite{us}. There we had a class of
actions with worldvolume local symmetries and rigid symmetries due to
isometries of the $adS$ space. After gauge fixing local symmetries, we found
that the gauge-fixed theory has conformal symmetry.
 The basic idea is the same in supergravity: the gauge-fixed action can be
constructed in background covariant gauges which respect the  symmetry of the
background. In the case of interest it is a conformal symmetry associated with
the $adS$ background.

Our general analysis applies to the following class of theories.
i)The so called Poincar\'{e} supergravities (ungauged) for which the $adS_{p+2}
\times S^{d-p-2-k}\times E_k$  background with some form-fields is a solution
of classical field equations.
ii) Any of the gauged supergravities for which
the $adS$ space is known to be a consistent solution of field equations. In
some
cases these gauged supergravities correspond to the massless modes of the
higher dimensional supergravity compactified on a sphere or on some other
compact space. This is the simplest case and for $adS_5$ it will be studied in
detail.

We  first remind the essential features of the background field method for the
supergravity in general and specify it for the case of a background with
isometries in Sec. 2. In Sec. 3 we derive the conformal symmetry of the
generating functional for the Green functions of the supergravity in $adS$
space in the bulk. In Sec. 4 we deduce the simplified rigid conformal symmetry
for correlators at the boundary of the $adS$ space. In Sec. 5 we suggest how to
compare our
results with available calculations. Finally  in the discussion we explain that
the  part
of the Maldacena's conjecture about conformal symmetry in supergravity is
proved. We explain why this is not  yet the proof of the exact correspondence
with the conformal Yang-Mills theory.

\section{Background field method in supergravity and isometries}

The background method is best described in condensed De Witt's notation
\cite{DeWitt} where the gauge symmetry of the classical action means that
\begin{equation}
S_{cl}[\Phi] = S_{ cl} [\Phi+ \delta \Phi] \ ,
\end{equation}
where $\Phi^i$ includes all fields of supergravity and
\begin{equation}
\delta_{ loc} \Phi^i = R^i{}_\alpha [\Phi] \xi^\alpha_{ loc}\ ,
\end{equation}
and $\xi^\alpha_{ loc}$ is the infinitesimal parameter of the local symmetry.
In presence of
a background field $\phi^i$, the fields of supergravity are shifted and the
classical action is given by $S_{cl}[\phi +\Phi]$. This action is invariant
under {\it two types of symmetries}.

1.  Transformations which affect only the quantum fields $\Phi$ which will
become the integration variables in the path integral:
\begin{equation}
\delta_{quant} \phi^i = 0 \qquad
\delta_{ quant} \Phi^i = R^i{}_\alpha [\phi +\Phi] \xi^\alpha_{ loc} \ .
\end{equation}
This results in
 \begin{equation}
\delta_{ loc}(\phi^i +  \Phi^i )= R^i{}_\alpha [\phi +\Phi] \xi^\alpha_{ loc} \
,
\end{equation}
which is a symmetry of the action $S_{cl}[\phi +\Phi]$.

2. Background symmetry. Transformations of the background fields $\phi$ are
accompanied
by transformations of the quantum fields $\Phi$:
\begin{equation}
\delta_{G} \phi^i = R^i{}_\alpha [\phi ] \xi^\alpha_{ loc}
 \qquad
\delta_{ G} \Phi^i =  R^i{}_{\alpha, j}  [\phi ]\Phi ^j   \xi^\alpha_{ loc} \ .
\label{back}
\end{equation}
This results in
 \begin{equation}
\delta_{ G}(\phi^i +  \Phi^i )= R^i{}_\alpha [\phi +\Phi] \xi^\alpha_{ loc} \ ,
\end{equation}
so that the action $S_{cl}[\phi +\Phi]$ is invariant.

In view of the local symmetries, to define the path integral we have to add to
the classical action of supergravity some gauge-fixing conditions and the ghost
actions. The generating functional of the Green functions in supergravity is
given by
\begin{equation}
\exp ^{iW[\phi, J]} = \int d\Phi d\bar c dc dc_3 \exp ^{i S [\phi,  \Phi, \bar
c, c, c_3, J] } \ .
\label{path}\end{equation}
Here $S $ consists of 4 terms, a classical action in the background, the
gauge-fixing  action, the ghost action and the source term.
\begin{equation}
 S[\phi,  \Phi, \bar c, c, J ] =  S_{cl}[\phi +\Phi] + S_{g.f} [\phi, \Phi]
+S_{ghost}[\phi, \Phi, \bar c, c, c_3] + J_i \Phi^i \ .
\label{integrand}\end{equation}
The integration in the path integral is performed over all  quantum
supergravity fields $\Phi^i$ and over the Faddeev-Popov anti-ghosts $\bar c$
and   ghosts $c$ and  over the third  ghost $c_3$. The term
$S_{g.f} [\phi, \Phi]$ is designed to break all local symmetries of the
classical action. In this way the local gauge symmetry of the first type,
$\delta_{quant}$  is broken and replaced by the BRST symmetry due to the
presence of ghosts.  However the second, the background  symmetry of the
classical
action, is preserved by both the gauge-fixing condition in background covariant
gauges,  by the ghost action
and by the source term.
\begin{equation}
 S[\phi,  \Phi, \bar c, c, J] =  S[\phi+\delta_{G} \phi ,  \Phi+ \delta_{ G}
\Phi, \bar c+ \delta_{ G}\bar  c,  c+ \delta_{ G} c, c_3+ \delta_{ G} c_3, J +
\delta_{G}J] \ .
\end{equation}
Since we are mostly interested here in the tree approximation of supergravity,
we will not provide here the background symmetry transformations of the
ghosts, it is sufficient to notice that it exists. However, the transformation
of source fields which preserves the background symmetry of the integrand in
the
path integral is important for our purpose and it is given below:
\begin{equation}
\delta_{G} J_i =  J_jR^j_{\alpha, i}[\phi]   \xi^\alpha \ .
\end{equation}
For example, the gravitational part of our integrand (\ref{integrand}) in the
path integral  (\ref{path}) is
\begin{eqnarray}
S&=& S_{cl} [g_{\mu\nu} + h_{\mu\nu}]+ S_{g.f} [g_{\mu\nu} , h_{\mu\nu}]
\nonumber\\
&+& S_{ghost}[g_{\mu\nu} , h_{\mu\nu},  \bar c, c]
+ \int d^d x \sqrt {g} \; h_{\mu\nu} J^{\mu\nu}
\end{eqnarray}
where $\sqrt {g}\equiv \sqrt {|det g_{\mu\nu}|}$ and
\begin{equation}
 S_{g.f} [g_{\mu\nu} , h_{\mu\nu}] = \int d^d x  \sqrt { g} \;  {1\over 2}
D^{\rho} h_{\rho\mu}\,g^{\mu\nu}D^\sigma h_{\sigma\nu}\ ,
\end{equation}
and $D^\mu$ is a covariant derivative in the background metric $g_{\mu\nu}$.
For the  vector fields we have
\begin{equation}
 S_{g.f} [g_{\mu\nu} , A_\mu] + S_{source}= \int d^d x  \sqrt { g} \;
{1\over 2} (D^{\mu} A_{\mu})^2 + \int d^d x \sqrt {g} \; A_{\mu} J^{\mu} \ .
\end{equation}

It is usual to consider  the gauge-fixing term quadratic in quantum fields, as
in the examples above, in DeWitt's notation this is
${1\over 2} \Phi^i  F_{ij}[\phi] \Phi^j$. Together with the part of the
classical action  quadratic in fields, $S_{cl}^2$, it gives
\begin{equation}
(S_{cl}^2 + S_{g.f}) [\phi, \Phi] = {1\over 2} \Phi^i (S_{,ij}[\phi] +
F_{ij}[\phi] )\Phi^j
\end{equation}
 This defines
the free background dependent Green functions $G^{ij}[\phi]$
\begin{equation}
-\left( S_{ij} + F_{ij}\right) [\phi] G^{jk}[\phi]= \delta_i{}^k \ .
\end{equation}
The basic property of the differential operators $(S_{ij} + F_{ij})[\phi]$ and
higher order vertices $S_{,ijk \dots}[\phi] $ in the background field method is
that they transform covariantly under the background field transformations.
This means that for each upper index the transformation is as in (\ref{back})
for $\Phi^i$ and for each  index down as for the source $J_i$.

Thus we have reminded here the well known property of the background field
method for supergravity which allows to construct the generating functional for
the Green functions of the theory in background covariant way, i.e. by keeping
the following symmetry:
\begin{equation}
W[\phi, J] = W[\phi + \delta_G \phi , J + \delta_G J] \ .
\end{equation}
This background symmetry is also present in the effective action of the theory:
\begin{equation}
\Gamma [\phi, \tilde \Phi] = W[\phi, J] - J_i \tilde \Phi^i \ ,
\end{equation}
where $J$ is replaced by the function of the background and classical
fields
according to the solution of
\begin{equation}
\frac{\delta W}{\delta J_i}[\phi,J(\phi,\tilde\Phi)]=\tilde\Phi^i\ .
\end{equation}
Here
one has to take into account that the background transformation of the field
$\tilde
\Phi^i$ is given by
\begin{equation}
\delta_{G} \tilde \Phi^i = R^i{}_{\alpha, j}  [\phi] \tilde \Phi^j \xi^\alpha \
{}.
\end{equation}
In case the background has isometries, which means that for some part
of $\xi^\alpha$, which will be denoted by $\xi^\alpha_{ K}$,
\begin{equation}
\delta_{G}^{ Killing} \phi^i = R^i{}_\alpha [\phi ] \xi^\alpha_{ K}=0 \ ,
\end{equation}
The generating functional for the Green functions of the theory has the
following symmetry\footnote{We assume here that the Jacobian of the background
transformations on the quantum fields is trivial, i.e. that the background
symmetry has no anomalies.}
\begin{equation}
W[\phi, J] = W[\phi , J + {\cal L}_{\xi_{ K}}
 J] \ .
\label{correl}\end{equation}
Here the transformation of the sources $J$ for the particular case we consider,
when the background has isometries, is reduced to an action of the Lie
derivative for  a given field with respect to the Killing vectors.
For the effective action we have the following symmetry:
\begin{equation}
\Gamma [\phi, \tilde \Phi ] = \Gamma [\phi, \tilde \Phi  + {\cal L}_{\xi_{ K}
}\tilde \Phi ] \ .
\label{effaction}\end{equation}
When the path integral of supergravity is expanded near the saddle point $\Phi
= \tilde \Phi$ , one gets in the first approximation the generating functional
for the correlators and the effective  action describing all tree diagrams.
\begin{equation}
W_{tree}[\phi, J] = S_{cl}[\phi +\tilde \Phi] + S_{g.f} [\phi, \tilde \Phi]   +
J_i \tilde \Phi^i\ .
\label{treecorrel}\end{equation}
\begin{equation}
\Gamma_{tree}  [\phi, \tilde \Phi ] = S_{cl}[\phi +\tilde \Phi] + S_{g.f}
[\phi, \tilde \Phi]
\label{treeeff}\end{equation}
\begin{equation}
-{ \delta \Gamma_{tree} [\phi, \tilde \Phi] \over \delta \tilde \Phi^i} = J_i\
{}.
\label{treefield}\end{equation}
The iterative solution of eq. (\ref{treefield}) gives $\tilde \Phi^i$ as a
functional of the sources. It involves background covariant free Green
functions $G^{ij}[\phi] $ and vertices $S_{,ijk}[\phi] , S_{,ijkl}[\phi], \dots
$ of the theory. Symbolically the tree solution for the field $\tilde \Phi^i $
can be written as
\begin{equation}
\tilde \Phi^i  =  \left( J_j +  {1\over 2}J_p G^{pm} S_{,mjn}[\phi]
J_qG^{qn}  + \dots  \right) G^{ji}[\phi]  \ .
\end{equation}
By inserting this tree field back into the r.h.s of eq. (\ref{treecorrel}) for
$W_{tree}$ one gets the generating functional of all connected Green functions
of the theory in the tree approximation.


\section{Conformal symmetry of  supergravities in $adS$ spaces in the bulk}

The new interesting feature of the  background field method comes from the fact
that when we consider supergravity in the fixed $adS$ background we are allowed
to consider  only those symmetries of the generating functional for the Green
functions or effective action  which do not change the background. The
background has isometries generated by the Killing vector fields $\xi_{K}$.
Therefore the symmetry of the generating functional is reduced to the action of
the Lie derivatives with respect to the Killing vectors of the $adS$ space.

The metric of our background  $adS_{p+2}
\times S^{d-p-2}$ geometry is
\begin{equation}\label{nearmetric1}
ds^2= \left( {r\over R}\right)^{2\over w} d {x}^m\eta_{mn}dx^n
+\left ({R\over r}\right)^2 [dr^2 + r^2d\Omega^2 ]\, .
\end{equation}
One can also  rewrite this metric  as ($m=0,\dots , p ; \; m' = p+1, \dots ,
d-1; \; r^2= y_{m'} y_{m'}$)
\begin{equation}\label{nearmetric2}
ds^2= \left( {r\over R}\right)^{2\over w} d {x}^m\eta_{mn}dx^n
+\left ({R\over r}\right)^2 dy^{m'}\delta_{m'n'}dy^{n'} \, .
\end{equation}
In these coordinates the infinitesimal action of the $SO(p+1,2)$ isometry group
is \cite{us}
\begin{eqnarray}
\label{conftrans}
\delta_{adS} (\xi)  x^m &=& - \hat \xi^m (x,r) = -\xi^m(x) - (wR)^2 \left(\frac
Rr\right)^{\ft2w} \Lambda_K^m\,,\nonumber \\
\delta_{adS}  (\xi) y^{m'}&= &- \hat \xi^{m'} (x,r) = w \Lambda_D(x)
y^{m'}\nonumber \\
\label{conftransr}
\delta_{adS}  (\xi)  r &=&- \hat \xi^r (x,r) =  w \Lambda_D(x) r \,,
\end{eqnarray}
where
\begin{eqnarray}
\xi^m (x) &=& a^m + \lambda_M^{mn} x_n + \lambda_D x^m + (x^2\Lambda_K^m - 2
x^m x\cdot \Lambda_K)\,, \nonumber\\
\Lambda_D (x) &=& \ft 1d \partial_m \xi^m = \lambda_D - 2 x\cdot
\Lambda_K
\label{rigid}\end{eqnarray}
and $a^m, \lambda_M^{mn}, \lambda_D, \Lambda_K^m$ are the constant parameters
associated with translations $P_m$, Lorentz transformations $M_{mn}$,
dilatations $D$ and special conformal transformations $K_m$.
 In case that $d> p+2$, there is also a sphere $S^{d-p-2}$ and we have
$SO(d-p-1)$ R-symmetry
\begin{equation}
\delta_{SO(d-p-1)} y^{m'} = \Lambda^{m'}{}_{n'} y^{n'}
\end{equation}
Now we can apply the $adS$ and the R-transformations to study the symmetry of
the generating functional of the supergravity  in the background covariant
gauges.

We will explain the main result for the simpler case of only $adS$ background
in the case of gauged supergravity. The generalization to the more general
situation when the sphere is present is technically more involved. Also for
definiteness we will focus on the case of $adS_5$ which describes the massless
part of ten-dimensional supergravity compactified on $adS_5\times S^5$. This
part of Maldacena's conjecture was developed in
\cite{FF,polyakov,witten,FFZ,Freedman} where actual calculations supporting the
conjecture were presented.

 The path integral of the 5-dimensional N=8 gauged supergravity in the $adS_5$
background is symmetric under the isometries generated by the Killing vector
fields, $\delta_{adS}(\xi) g_{\mu\nu} = 0$
and we will use the notation $\mu =0,1,2,3,r$. Here $m=0,1,2,3$ are coordinates
of the 4-dimensional boundary which is at  $r\rightarrow \infty$.

First we focus on supergravity in the bulk.
The action of $adS$ symmetry on all sources to the supergravity fields is
generated by the Lie derivative with respect to the Killing vectors $\xi$.
These in turn coincide with what is known in supergravity literature as
`general covariance' transformation of various fields in the action. By
combining these two notions we are getting the following symmetry
transformations for the sources of the gravitational, vector and scalar fields,
$J^{\mu\nu}, J^\mu, J$ which are coupled to the supergravity fields. The
generating functional for the Green functions (\ref{correl}) is symmetric under
the following transformations of sources:
\begin{equation}
-{\cal L}_{\xi} J^{\mu\nu} = \hat \xi^{\lambda} \partial_{\lambda}  J^{\mu\nu}
-
\partial_\lambda  \hat \xi^\mu  J^{\lambda\nu} - \partial_\lambda  \hat \xi^\nu
 J^{\mu\lambda}
\label{grav}
\end{equation}
\begin{equation}
-{\cal L}_{\xi}  J^{\mu} = \hat \xi^{\lambda} \partial_{\lambda}  J^{\mu} -
\partial_\lambda  \hat \xi^\mu  J^{\lambda } \label{vec}
\end{equation}
\begin{equation}
-{\cal L}_{\xi}  J = \hat \xi^{\lambda} \partial_{\lambda} J \label{scal}
\end{equation}
and the corresponding transformations on fermionic sources. Here $\hat
\xi^{\lambda} = \hat \xi^{m}$ for $\lambda=0,1,2,3$  is defined in eq.
(\ref{conftrans}) and the last one, the $r$-component is  $\hat \xi^r$ is equal
to $ - w \Lambda_D(x) r$ as one can see from eq.
 (\ref{conftransr}).

The effective action (\ref{effaction}) is symmetric under the following
transformations of the effective fields
\begin{equation}
-{\cal L}_{\xi} \tilde h_{\mu\nu} = \hat \xi^{\lambda} \partial_{\lambda}
\tilde h_{\mu\nu} + \partial_\mu  \hat \xi^\lambda \tilde h_{\lambda\nu} +
\partial_\nu  \hat \xi^\lambda  \tilde h_{\mu\lambda}
\label{grav1}
\end{equation}
\begin{equation}
-{\cal L}_{\xi}  \tilde A_{\mu} = \hat \xi^{\lambda} \partial_{\lambda}
\tilde
A_{\mu}  + \partial_\mu  \hat \xi^\lambda   \tilde A_{\lambda}  \label{vec1}
\end{equation}
\begin{equation}
-{\cal L}_{\xi}  \tilde \Phi = \hat \xi^{\lambda} \partial_{\lambda}  \tilde
\Phi \label{scal1}
\end{equation}
and the corresponding transformations on fermionic fields.

The transformations shown in eqs. (\ref{grav}), (\ref{vec}),
 (\ref{scal}) represent the conformal symmetry of the generating functional for
the correlators  of the supergravity fields in the bulk at finite values of
$r$.  Note that so far our sources are defined in the bulk and differentiating
the generating functional over the sources one can get all correlators of
supergravity fields in the bulk.

If the actual calculation of the generating functional is performed, e. g. the
term quadratic in $J$ is found it takes the form
\begin{eqnarray}
&&W(g_{\mu\nu}(r) , J^{\mu\nu} (x,r) , J^\mu (x,r), J(x,r),...) \nonumber\\
 \nonumber\\
&&= {1\over 2} \int d^4 x dr d^4 x' dr' J(x,r) G(x,r; x', r') J(x', r') + ...
\end{eqnarray}
This expression has to be invariant under the transformations (\ref{scal})
which puts constraints on the Green function $G$ in the bulk. The same takes
place for any other correlator.

\section{Conformal symmetry of supergravities on the boundary of the  $adS$
spaces}

The basic  idea developing Maldacena's conjecture \cite{maldacena} was
suggested by Gubser, Klebanov and  Polyakov \cite{polyakov}, and
Witten\cite{witten}. It  was to place the sources to the supergravity fields at
the boundary and calculate the correlators of the fields on the boundary and
compare them with those of the YM theory.

In our context the advantage of considering  the sources to the supergravity
fields located only at the boundary of the $adS$ space is the dramatic
simplification of the  conformal symmetry comparative to the one in the bulk.

To find these symmetries we have to consider the limit of our
conformal symmetries in the bulk to the boundary at $r\rightarrow \infty$. We
specify the case with $adS_5, \; w=1$.
We will first change the variables to $z={R^2\over r}$ in which the $adS$ part
of the metric is conformally flat $ds^2= {R^2\over z^2} (dx_m^2 + dz^2)$. In
these variables the boundary is at $z\rightarrow 0$.
Clearly,
\begin{equation}
\label{limit}
  \hat \xi^m (x,z)_{z\rightarrow 0}  = (\xi^m(x) + z^2  \Lambda_K^m)_{
z\rightarrow 0} \Longrightarrow \xi^m(x) \, \end{equation}
and
\begin{equation}
( \hat \xi^ z )_{ z\rightarrow 0 }  \Longrightarrow   \Lambda_D(x) z \,,
\label{r}\end{equation}
where the parameters of the rigid conformal symmetry $\xi^m (x) $ and
$\Lambda_D (x)$ are given in eq. (\ref{rigid}). We will denote the sources
placed at the boundary, which are $z$-independent by ${\cal J}$.
To find the boundary limit of the conformal transformations from the bulk we
note that on $z$-independent functions
\begin{equation}
\left( \hat \xi^m (x,z) \partial _m + \hat \xi^m (x,z) \partial_z \right) _{
z\rightarrow 0}  \Longrightarrow  \xi^m (x) \partial _m \ .
\end{equation}
To study the limit to the boundary on tensors we need to use
\begin{eqnarray}
&&\left(\partial_m \hat \xi^n (x,z) \right) _{ z\rightarrow 0} \Longrightarrow
\partial_m \xi^n (x)\\
&& \left(\partial_z \hat \xi^z (x,z) \right) _{
z\rightarrow 0}  \Longrightarrow  \Lambda_D (x) \\
&&\left(\partial_z \hat \xi^n (x,z) \right) _{ z\rightarrow 0} \Longrightarrow
2  \Lambda_K^n (x) z \Longrightarrow 0 \\
&&\left(\partial_m \hat \xi^z (x,z) \right) _{ z\rightarrow 0}  \Longrightarrow
\partial_m \Lambda_D (x) z \Longrightarrow 0
\end{eqnarray}

It follows that the components of the contravariant tensors in the bulk
direction $z$ are not mixed anymore with those in directions $x$.

\begin{equation}
-{\cal L}^{bound}_{\xi} {\cal J}^{mn} =  \xi^{l}(x) \partial_{l}  {\cal J}^{mn}
- \partial_l   \xi^m  {\cal J}^{ln} - \partial_l  \xi^n  {\cal J}^{m l}
\label{boundgrav}
\end{equation}
\begin{equation}
-{\cal L}^{bound} _{\xi}  {\cal J}^{m} = \xi^{l} \partial_{l}  {\cal J}^{m} -
\partial_l  \xi^m  {\cal J}^{l} \label{boundvec}
\end{equation}
\begin{equation}
-{\cal L}^{bound} _{\xi}  {\cal J} = \xi^{l} \partial_{l} {\cal J}
\label{boundscal}
\end{equation}
with $\xi^m(x) $ defined in eq. (\ref{rigid}).

The transformation of the remaining components of the sources is
\begin{equation}
-{\cal L}^{bound}_{\xi} {\cal J}^{mz} =  \xi^{l}(x) \partial_{l}  {\cal J}^{mz}
- \partial_l   \xi^m  {\cal J}^{lz} - \Lambda_D(x) {\cal J}^{m z}
\label{mz}
\end{equation}
\begin{equation}
-{\cal L}^{bound}_{\xi} {\cal J}^{zz} =  \xi^{l}(x) \partial_{l}  {\cal J}^{zz}
 - 2 \Lambda_D(x) {\cal J}^{m z}
\end{equation}
\begin{equation}
-{\cal L}^{bound} _{\xi}  {\cal J}^{z} = \xi^{l} \partial_{l}  {\cal J}^{z}
-\Lambda_D(x)  {\cal J}^{z} \ .\label{z}
\end{equation}

The generating functional depending on boundary sources has the following
symmetry:
\begin{equation}
W^{bound} [ {\cal J}  ] = W [{\cal J}  + {\cal L}^{bound}_{\xi}
 {\cal J}]
\label{conf}\end{equation}
where the transformation of all sources at the boundary are given in eqs.
(\ref{boundgrav})-(\ref{z}).

Note that we have suppressed here the dependence on internal indices of the
vectors and scalars since for our simple example  conformal symmetries do not
act on them, the only relevant ones are those which show the behavior under
general coordinate transformations.

\section{Comparison with available calculations}
Consider here the calculations of the 2-points and 3-points correlators
available in the literature \cite{polyakov,witten,Freedman}. For massless
scalars in Euclidean signature we have
\begin{equation}
W^{(2)} [{\cal J} ] = c\int d^4 x d^4 y {{\cal J}   (x) {\cal J} (y) \over
|x-y|^8}
\end{equation}
This answer is in agreement with our form of conformal symmetry which states
that under the transformations (\ref{boundscal}) the functional $W$ has to be
symmetric. These transformations include  translation,  dilatation, and Lorentz
transformations which are all obvious but also the special conformal
transformations, which are not so obvious.
The transformation of $W^{(2)}  [{\cal J} ]$
is
\begin{eqnarray}
\delta_\xi W^{(2)}   [{\cal J} ]=- c\int d^4 x \hspace{-6mm}&&d^4 y {1 \over
|x-y|^8}\left[
\left( {\cal L}_\xi {\cal J}   (x)\right)  {\cal J} (y) +
 {\cal J}   (x) \left( {\cal L}_\xi {\cal J} (y) \right)  \right]
 \nonumber\\
 = c\int d^4 x \hspace{-6mm}&&d^4 y {1 \over |x-y|^8}\nonumber\\ &&\left[
\left(\xi(x)^m\partial_m {\cal J}   (x)\right)  {\cal J} (y) +
 {\cal J}   (x) \left( \xi(y)^m\partial_m {\cal J} (y) \right)  \right]
\end{eqnarray}
After integration by parts, and using \eqn{rigid}, we obtain
\begin{eqnarray}
\delta_\xi W^{(2)}   [{\cal J} ]= c\int d^4 x\hspace{-6mm}&& d^4 y
{1 \over |x-y|^8} {\cal J}   (x)  {\cal J} (y) \\ &&
\left[-4\Lambda_D(x)-4\Lambda_D(y)+  8 \frac{(\xi(x)^m-\xi(y)^m) (x_m-y_m)}
{|x-y|^2}\right] \ .     \nonumber
\end{eqnarray}
Inserting the explicit expression of $\xi^m$ in \eqn{rigid} leads to
the cancellation of the terms in the square brackets.

For vectors in \cite{witten,Freedman} the relevant correlators are given for
2-point functions and in \cite{Freedman} for the 3-point functions. Instead of
doing a direct verification of our symmetries when ${\cal J}^m$ sources are
non-vanishing, as shown for scalars above we may use some  important
properties of the correlators established
in  \cite{Freedman} which will allow us to confirm that there is an agreement
with our form of symmetries.
In  \cite{Freedman} the boundary correlators  are found to be covariant under
inversion which means that the generating functional $W[{\cal J}^m]$ is
invariant under inversion. The relevant non-local functional  is also invariant
under translational symmetry. Therefore by applying inversion, translation and
another inversion one can find out that the generating functional is indeed
invariant under conformal symmetry as predicted by our analysis.
The functional  for the 2-point correlators of vector fields found in
\cite{witten,Freedman} has the form
\bea
\label{2point}
W^2 [{\cal J} ^m] = c\int d^4 x d^4 y {\cal J}_{a }^m (x) {\cal J}_{b }^n (y)
\delta^{ab}  \left( \partial_l\partial^l
\delta_{mn}-\partial_m\partial_n\right)  { 1\over |x-y|^4} \ .
\eea
The $f^{abc}$ part of the  3-point correlator  was found in \cite{Freedman}
\bea
\label{3point}
W^3 [{\cal J}^m] = &&c\int d^4 x d^4 y d^4 z {\cal J}_{a }^m (x) {\cal J}_{b
}^n (y)
{\cal J}_{c }^l f^{abc}\\
&& (k_1 D_{mnl}^{\rm
sym}(x,y,z) + k_2 C_{mnl}^{\rm sym}(x,y,z)),
 \eea
where the tensors $D$ and $C$ have particular form presented in
\cite{Freedman}.
It may be difficult but not impossible to establish that the generating
functional for the correlators  of supergravity fields on the boundary
calculated in \cite{Freedman} indeed has the required symmetry properties
required by  eqs. (\ref{conf}), (\ref{boundvec}). This would confirm  that the
choice of the inversion-covariant bulk-to-boundary Green functions in
\cite{Freedman}  corresponds to the correct choice of the background-field
covariant gauge in $adS$ background.

\section{Discussion}

In this note we have addressed the problem which naturally comes to the mind
of anybody familiar with the difference between  ungauged supergravity, gauged
supergravity and conformal supergravity. As phrased by K. Stelle: why should
Poincar\'{e} supergravity know anything about  conformal symmetry at all? The
answer is that in general, indeed, only conformal supergravity has local
conformal symmetry and both ungauged and gauged supergravities do not have
anything close to conformal symmetry. However, when placed in the consistent
$adS$ background with symmetries isomorphic to conformal group these two
versions of supergravity upon gauge-fixing do have conformal symmetry as
conjectured in \cite{maldacena}.
In particular the generating functional for all correlators of supergravity
fields everywhere in $adS$ space-time has conformal symmetries of the type
which were found previously  on the worldvolumes \cite{us} of brane actions.
They are characterized by  an  unusual form of the special conformal symmetry.
Now we have found the analogous symmetries in the space-time  supergravity.

We also studied a particular case when the sources of the supergravity fields
are placed only on the boundary of the $adS$ space-time, as proposed in
\cite{polyakov, witten}.  The conformal symmetry of supergravity in the bulk is
simplified and reduced to the simple rigid conformal symmetry, consisting as
usual of translations, Lorentz transformations, dilatations and special
conformal transformations. For example we found that the generating functional
of the correlators of scalar fields of supergravity on the boundary is
symmetric under the transformations of the  $z$-independent scalar source
${\cal J}$ placed at the boundary $z\rightarrow 0$:
\begin{equation}
\delta {\cal J} (x) = \left (a^m + \lambda_M^{mn} x_n + \lambda_D x^m +
(x^2\Lambda_K^m - 2
x^m x\cdot \Lambda_K)\right ) {\partial \over \partial x^m} \;   {\cal J}(x)
\end{equation}

It should also be stressed that in Witten diagrams \cite{witten,Freedman}, as
our analysis shows, the Green functions which are not touching the boundary are
in fact bulk-to-bulk Green functions which are different from bulk-to-boundary
correlators used in the calculations performed so far. Such bulk-to-bulk
correlators will be always present in the tree level boundary supergravity
starting with 4-point correlators as well as in all loop diagrams. These Green
functions are conformal covariant in the bulk and not only at the boundary, as
well as all vertices of the theory.

Thus we have found  a clear explanation via a background field method  why in
ungauged or gauged supergravity one encounters rigid conformal symmetry.
Note that our proof never used the tree approximation to the supergravity
correlators, it is correct for any loop approximation under condition that the
$adS$ background remains a consistent solutions of equations of motion with
quantum corrections. It will be interesting to establish whether this is true.

The observation made here does not indicate yet that the results of
calculations of correlators in tree level supergravity on the boundary have to
be the same as the one in large $N$ limit of ${\cal N}=4$ supersymmetric
Yang-Mills theory. For this to happen one still have to find out whether the
conformal correlators with rigid ${\cal N}=4$ supersymmetry
are unique. In general this is by no means clear, the correlators may depend on
few conformally invariant functions \cite{HoweWest} and the constants between
them may not be always fixed by the superconformal symmetry alone: calculations
in specific theories may be required to supply such coefficients\footnote{We
are grateful to P. West for clarification of  this issue.}. If superconformal
symmetry is not fixing the correlators in a unique way, some other reason for
precise conjectured agreement between supergravity and Yang-Mills theory has to
be found.

In conclusion, we have shown a surprisingly simple picture as to how a
conformal (and superconformal) symmetry is present in supergravities in $adS$
spaces. This part of the famous CFT/AdS correspondence \cite{maldacena} is no
conjecture anymore: if one uses the correct Feynman rules for the calculations
of tree diagrams in supergravity in the particular set up, explained in this
paper,
conformal symmetry is guaranteed to be there, in the form of eq. (\ref{correl})
in the bulk and (\ref{conf}) at the boundary.
\vskip 1 cm

{\it Note added} \,  After this paper was written we have seen the recent paper
of Liu and Tseytlin \cite{LiuTsey} where new calculations of supergravity
correlators on the boundary of the $adS$ space
are performed. In particular the aim was to find the graviton-dilaton-dilaton
correlator `without making a priory assumption about the conformal invariance
of the result'. A detailed study of these set of calculations (and any new one
which may appear soon) would be interesting to carry out from the perspective
of our proof of the conformal symmetry of the full generating functional for
the boundary correlators of supergravity on $adS$.

 \vskip 1 cm

\vskip 1 cm

We are grateful to the organizers and the participants of the `Super Five
Branes
and Physics in 5+1 dimensions' conference in Trieste in April 1998 for the
stimulating discussions of the preliminary form of the result presented in this
paper. R. K. also would like to acknowledge particularly useful conversations
related to this work with A. Matusis, A. Rajaraman and  R. Siebelink.
The work  of R. K. is supported by the NSF grant PHY-9219345, and the
work of A.V.P. is supported by
the European Commission TMR programme ERBFMRX-CT96-0045.

\end{document}